\documentclass[]{aastex631}
\usepackage{xcolor}
\usepackage{ulem}
\def\r{\textcolor{red}}
\def\b{\textcolor{blue}}
\def\c{\textcolor{cyan}}
\def\t{\textcolor{teal}}
\def\o{\textcolor{orange}}
\def\p{\textcolor{purple}}

\shorttitle{Blazar broad emission lines study}
\shortauthors{Xiao et al.}

\begin{document}

\title{An extensive study of blazar broad emission line: `Changing-look' blazars and `Baldwin effect'}

\correspondingauthor{Hubing Xiao}
\email{hubing.xiao@shnu.edu.cn}
\correspondingauthor{Junhui Fan}
\email{fjh@gzhu.edu.cn}

\author[0000-0001-8244-1229]{Hubing Xiao}
\affiliation{Shanghai Key Lab for Astrophysics, Shanghai Normal University, Shanghai, 200234, China}

\author{Junhui Fan}
\affiliation{Center for Astrophysics, Guangzhou University, Guangzhou 510006, China}
\affiliation{Key Laboratory for Astronomical Observation and Technology of Guangzhou, Guangzhou, 510006, China}
\affiliation{Astronomy Science and Technology Research Laboratory of Department of Education of Guangdong Province, \\
Guangzhou, 510006, China}

\author{Zhihao Ouyang}
\affiliation{School of Physics and Materials Science, Guangzhou University \\
Guangzhou, 510006, China}

\author{Liangjun Hu}
\affiliation{Wuhu No.27 Middle School, Wuhu, 241001, China}

\author{Guohai Chen}
\affiliation{Center for Astrophysics, Guangzhou University, Guangzhou 510006, China}
\affiliation{Key Laboratory for Astronomical Observation and Technology of Guangzhou, Guangzhou, 510006, China}
\affiliation{Astronomy Science and Technology Research Laboratory of Department of Education of Guangdong Province, \\
Guangzhou, 510006, China}

\author{Liping Fu}
\affiliation{Shanghai Key Lab for Astrophysics, Shanghai Normal University, Shanghai, 200234, China}

\author{Shaohua Zhang}
\affiliation{Shanghai Key Lab for Astrophysics, Shanghai Normal University, Shanghai, 200234, China}

\begin{abstract}
It is known that the blazar jet emissions are dominated by non-thermal radiation while the accretion disk jets are normally dominated by thermal emission. 
In this work, our aim is to study the connection between the two types of emission by investigating the correlation between the blazar emission line intensity property, which embodies the nature of accretion disk, and the $\gamma$-ray flux property, which is the representative of jet emission.
We compiled a sample of 656 blazars with available emission line equivalent widths ($EW$), the GeV $\gamma$-ray flux, and the SED information from the literature.
In this work, we found 55 previous BCUs are now identified as FSRQs, and found 52 `Changing-look’ blazars based on their $EW$ and 45 of them are newly confirmed.
These `Changing-look’ blazars have a larger accretion ratio (${\dot M}/{\dot M}_{\rm Edd}$) than BL Lac objects.
In addition, we suggest that the lower synchrotron peak blazars (LSPs) could be the source of `Changing-look’ blazars because 90.7\% of the `Changing-look’ blazars in this work are confirmed as LSPs. 
An anti-correlation between $EW$ and continuum intensity, the so-called global `Baldwin effect’ (BEff) has been confirmed. 
We suggest the steeper global BEff observed for blazar than for radio-quiet active galactic nuclei (RQ-AGNs) is caused by the inverse Compton scattering of broad-emission-line photons. 
This interpretation is further supported by the positive correlation between the emission line $EW$ and intrinsic inverse Compton luminosity.
\end{abstract}

\keywords{}

\section{Introduction}
Blazars, one of the most extreme subclasses of active galactic nuclei (AGNs), show extreme observational properties, such as rapid and strong multi-wavelength variability, high and variable polarization, strong and variable $\gamma$-ray emissions and apparent superluminal motion at radio frequencies \citep{Wills1992, Urry1995, Fan2002, Fan2014, Fan2021, Villata2006, Xiao2015, Gupta2016, Thorn2018, Lister2018, Xiao2019, Abdollahi2020}.
These observational properties are characteristic of the relativistic jet which points toward the observer.
Blazar emit radiation, which is non-thermal dominated, across the entire electromagnetic spectrum.
A typical blazar broadband spectral energy distribution (SED) displays a two-bump structure.
The lower energy bump peaks in the range of infrared to X-ray and is attributed to the synchrotron radiation of relativistic electrons, while the origin of the higher energy bump which peaks at X-ray to $\gamma$-ray wavelengths is under debate.
There are two models for the higher-energy bump, the leptonic model states that the higher energy bump is attributed to the inverse Compton (IC) scattering \citep{Blandford1979, Sikora1994, Sokolov2005, Abramowski2015, Zheng2016, Tan2020}, while the hadronic model interprets that the higher-energy bump is attributed to the proton synchrotron radiation and secondary particle cascade \citep{Mucke2001, Dimitrakoudis2012, Zheng2013, Diltz2015, IceCube2018, Xue2021}.

Both the optical spectrum and SED are used to make classifications for blazars.
Historically, blazars are divided into two classes, namely flat spectrum radio quasars (FSRQs) and BL Lacertae objects (BL Lacs), according to their optical spectra.
The former is characterized by strong emission lines with the rest-frame equivalent width ($EW$) of the strongest emission line greater than $5\, \mathring{\rm A}$, while the latter one shows featureless optical spectrum or weak emission lines with the $EW$ of the strongest emission line less than $5\, \mathring{\rm A}$ \citep{Stocke1990, Stocke1991, Stickel1991, Urry1995, Scarpa1997}.
The separation value of $5\, \mathring{\rm A}$ was deduced from an examination of the line strength of a representative sample of FSRQs from the Parkes catalogue \citep{Wilkes1986}, using the same limit in the selection of the X-ray selected BL Lacs (XBLs) from the \textit{Einstein} Extended Medium Sensitivity Survey (EMSS) \citep{Stocke1990}.
This separation value was also applied in \citet{Stickel1991} to select BL Lacs from $1\, {\rm Jy}$ catalogue of radio sources \citep{Kuhr1981}, and 34 BL Lacs were obtained.
It is clear that the separation value of $5\, \mathring{\rm A}$ is rather arbitrarily settled.
A Doppler boosted non-thermal continuum could swamp out spectral emission lines \citep{Blandford1978, Xiong2014}, and $EW$ greater than $5\, \mathring{\rm A}$ may be the result of a particular low-state of jet activity.

The classification types, which based on the optical spectrum of AGNs and that of blazars were harmonious until the discovery of `Changing-look' AGNs \citep{Matt2003, Bianchi2005}.
The shifting between Type I AGNs and Type II AGNs, or the shifting between FSRQs and BL Lacs has brought great challenges to the AGN unification model \citep{Marchese2012, Shappee2014, Isler2013, Mishra2021, Pena2021}.
There are explanations for the shifting, among them the sudden change of accretion ratio arising the shift seems promising.

The SED features, e.g. synchrotron peak frequency ($\nu_{\rm s}$) and Compton dominance are employed to make classifications for blazars in previous works.
A parabolic function is widely used to describe the blazar SEDs in the diagram of ${\rm log}\, \nu F_{\nu} - {\rm log}\, \nu$ since it was proposed in \citet{Landau1986}.
\citet{Nieppola2006} fitted SEDs in the form of ${\rm log}\, \nu F_{\nu} - {\rm log}\, \nu$ with parabolic function for 308 blazars and classified BL Lacs into low synchrotron peak BL Lacs (LBLs, ${\rm log}\, \nu_{\rm s} < 14.5$), intermediate synchrotron peak BL Lacs (IBLs, $14.5 < {\rm log}\, \nu_{\rm s} < 16.5$), and high synchrotron peak BL Lacs (HBLs, ${\rm log}\, \nu_{\rm s} > 16.5$) based on the synchrotron peak frequency.
Similarly, \citet{Fan2016} calculated SEDs by fitting the multi-wavelength data with parabolic function for a larger sample of 1392 blazars and classified blazars into low synchrotron peak sources (LSPs, ${\rm log}\, \nu_{\rm s} \le 14.0$), intermediate synchrotron peak sources (ISPs, $14.0 < {\rm log}\, \nu_{\rm s} \le 15.3$), and high synchrotron peak sources (HSPs, ${\rm log}\, \nu_{\rm s} > 15.3$).
Recently, \citet{Paliya2021} proposed that Compton dominance (CD) can be considered as one such parameter to reveal the physics of the non-thermal jets in beamed AGNs, suggesting broad-emission-line blazars being more Compton-dominated sources, and separated blazars as HCD (high ${\rm CD} > 1$) and LCD (low ${\rm CD} < 1$).

It is no doubt that the last decade has been a golden age for blazar research in high energy bands due to the all-sky survey carried out by the Large Area Telescope (\textit{Fermi}-LAT, \citealt{Atwood2009}).
The fourth catalogue of active galactic nuclei (AGNs) detected by the \textit{Fermi}-LAT (4LAC, \citealt{Ajello2020}) between 2008 August 4 and 2016 August 2 contains 3511 sources in 4LAC\_DR2, among them 3437 blazars are included.
The blazar $\gamma$-ray emission from a non-thermal mechanism dominates the entire electromagnetic radiation \citep{Ghisellini2011, Ghisellini2014, Xiong2014}.

In this work, we investigate the optical emission line strength and $\gamma$-ray intensity (including the inverse Compton intensity) to study the thermal emission and its connection with the non-thermal emission in blazars.
In section 2, we define our sample, and our results will be presented in section 3.
The discussions will be given in section 4.
Section 5 presents our conclusions.

\section{Sample}
We collect blazar emission-line profiles from \citet{Paliya2021}, in which 674 \textit{Fermi} sources are included. 
According to 4LAC\_DR2 \citep{Ajello2020}, 17 out of 674 blazars in \citet{Paliya2021} are not considered as blazars.
Besides, 4FGL J0014.1+1910 is excluded because it shows a noisy spectrum and gives no $EW$ of emission lines.
At last, we have a sample of 656 \textit{Fermi} blazars with emission line $EW$ and redshift from \citet{Paliya2021}, $\gamma$-ray intensity and SED features from 4LAC\_DR2 \citep{Ajello2020}.
We list our sample and the parameters in Table \ref{tab_og}. 
Among the 656 sources, 55 sources are classified as BCUs (blazar candidates of uncertain type), 51 sources are classified as BL Lacs, and 550 sources are classified as FSRQs according to the classification types in 4LAC\_DR2.

\section{Results}
\subsection{The blazars classification}
The redshift distributes from 0.027 to 4.314 with a mean value of $1.163 \pm 0.665$ for the blazars in our sample.
There are 47 sources with detected ${\rm H{\alpha}}$ emission line, the average equivalent width of ${\rm H{\alpha}}$ is $\langle EW_{\rm H{\alpha}} \, (\mathring{\rm A})\rangle =258.53 \pm 59.21$;
160 sources with detected $\rm H{\beta}$ emission line, and the $\langle EW_{\rm H{\beta}} \, (\mathring{\rm A}) \rangle =121.82 \pm 13.61$;
482 sources with detected MgII emission line, and the $\langle EW_{\rm MgII} \, (\mathring{\rm A}) \rangle =79.72 \pm 9.50$;
193 sources with detected CIV emission line, and the $\langle EW_{\rm CIV} \, (\mathring{\rm A}) \rangle =106.63 \pm 9.41$;

We are able to determine the blazar type for these 55 BCUs based on their optical spectra. 
All of the 55 BCUs show $EW$ of emission lines greater than 5 $\mathring{\rm A}$, thus we suggest these sources are FSRQs and list them in Table \ref{tab_u}.
Based on the classic blazar type criterion of 5 $\mathring{\rm A}$, we find 52 blazars that show type change, denote as `Changing-look' blazars, including 10 FSRQs change their types to BL Lacs (`F$\to$B') and 42 BL Lacs change their types to FSRQs (`B$\to$F'), and list them in Table \ref{tab_cl}.

\subsection{Correlations between equivalent width and continuum luminosity}
\begin{figure}[htbp]
\centering
\includegraphics[scale=0.9]{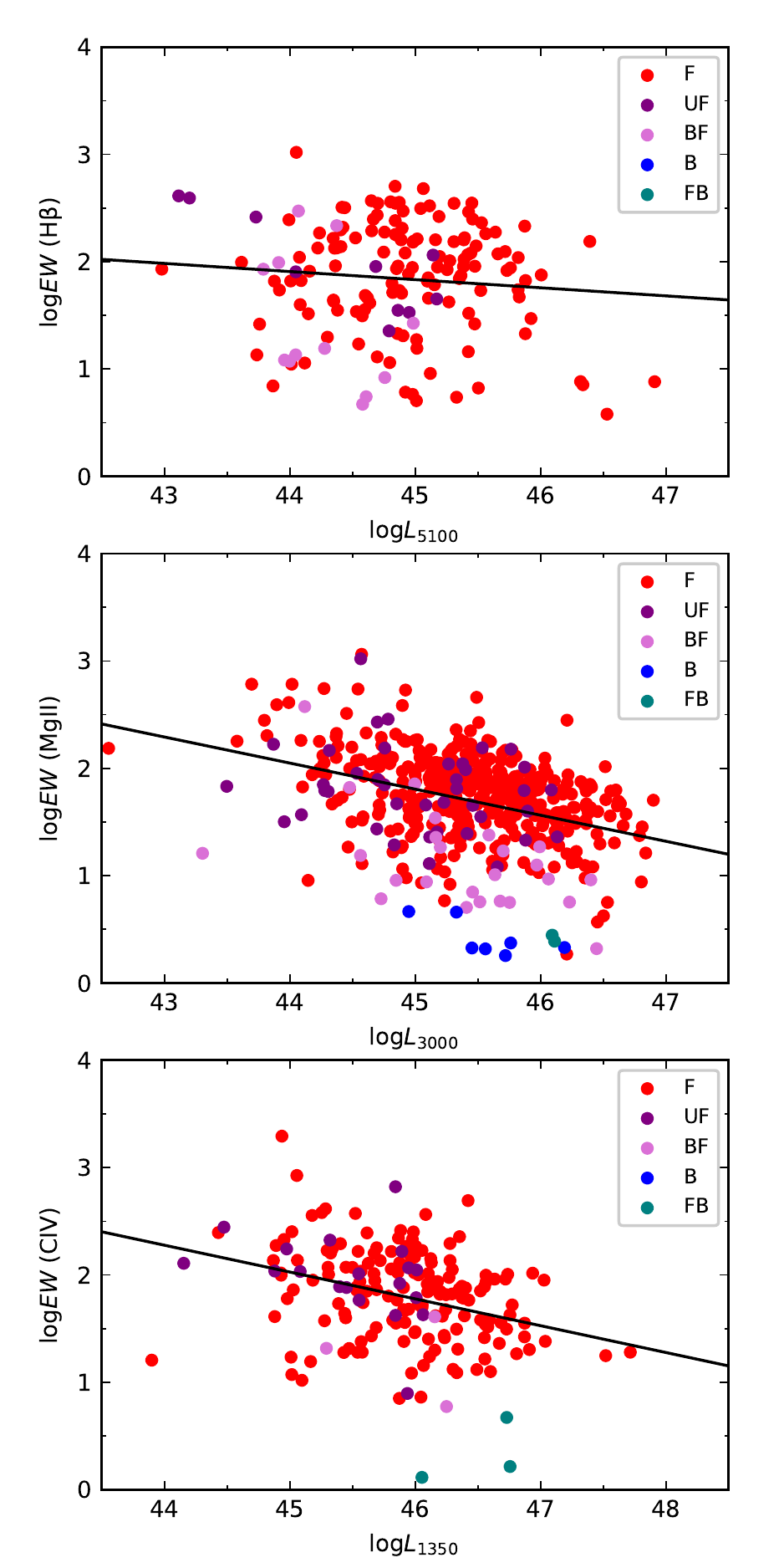}
\caption{The correlations between emission line ($\rm H{\beta}$, $\rm Mg{II}$, and $\rm C{IV}$) EW and corresponding optical continuum luminosity.
`F' stands for FSRQs, `UF' stands for the new confirmed FSRQs from BCUs in this work, `BF' stands for the `changing look' blazar that changes from BL Lac to FSRQs, `B' stands for BL Lacs, `FB' stands for the `changing look' blazar that change from FSRQs to BL Lacs.}
\label{Fig_cor_line}
\end{figure}

We collect the continuum luminosity ($L_{\rm \lambda}$) at $5100\,\mathring{\rm A}$, at $3000\,\mathring{\rm A}$, and at $1350\,\mathring{\rm A}$ from \citet{Paliya2021}, in which these continuum luminosities were calculated via empirical relations with emission line luminosities \citep{Shen2011, Shaw2012}:
\begin{equation}
{\rm log} \, L_{\rm 3000} = (1.016 \pm 0.003)\, {\rm log} \, L_{\rm MgII} + (1.22 \pm 0.11)
\end{equation}
\begin{equation}
{\rm log} \, L_{\rm 1350} = (0.863 \pm 0.009)\, {\rm log} \, L_{\rm CIV} + (7.66 \pm 0.41)
\end{equation}
\begin{equation}
{\rm log} \, L_{\rm 5100} = (0.802 \pm 0.049)\, {\rm log} \, L_{\rm H \beta} + (1.574 \pm 0.060)
\end{equation}
Figure \ref{Fig_cor_line} shows the correlations between the emission line $EW$ and the line continuum luminosity.
The linear regression results are illustrated in Table \ref{lin_cor_line}, where the linear correlation is expressed as $y = (a \pm \Delta a)x + (b \pm \Delta b)$, $N$ is the size of the considered sample, $r$ is a correlation coefficient, and $p$ is a chance probability.

\subsection{Correlations between equivalent width and GeV $\gamma$-ray parameters}
Assuming the GeV $\gamma$-ray photons follow a power law function and is expressed as
\begin{equation}
{\frac{dN}{dE}} = N_{\rm 0} E^{-\alpha_{\rm ph}},
\end{equation}
where $\alpha_{\rm{ph}}$ is the photon spectral index, and $N_{\rm 0}$ can be expressed as
$N_{\rm 0} = N_{(E_{\rm L}\sim E_{\rm U})}({\frac{1}{E_{\rm L}}-\frac{1}{E_{\rm U}}}),$ 
if $\alpha_{\rm ph}=2$, otherwise
$N_{\rm 0} = \frac{N_{(E_{\rm L}\sim E_{\rm U})}(1-\alpha_{\rm ph})}{(E_{\rm U}^{1-\alpha_{\rm ph}}-E_{\rm L}^{1-\alpha_{\rm ph}})},$
where $N_{(E_{\rm L}\sim E_{\rm U})}$ is the integral photons in units of ${\rm photons \cdot cm^{-2}\cdot s^{-1}}$ in the energy range of $E_{\rm L}$ - $E_{\rm U}$, where $E_{\rm L}$ and $E_{\rm U}$ correspond to 1 GeV and 100 GeV respectively.
The integral flux, $F$, in units of ${\rm GeV \cdot cm^{-2}\cdot s^{-1}}$, can be expressed in the form \citep{Fan2013, Xiao2015}
\begin{equation}
F = N_{(E_{\rm L}\sim E_{\rm U})}{\frac{E_{\rm U} \times E_{\rm L}}{E_{\rm U} - E_{\rm L}}}\ln \frac{E_{\rm U}}{E_{\rm L}}
\end{equation}
for $\alpha_{\rm ph}=2$, otherwise
\begin{equation}
F = N_{(E_{\rm L}\sim E_{\rm U})}{\frac{1-\alpha_{\rm ph}}{2-\alpha_{\rm ph}}}{\frac{(E_{\rm U}^{2-\alpha_{\rm ph}}-E_{\rm L}^{2-\alpha_{\rm ph}})}{(E_{\rm U}^{1-\alpha_{\rm ph}}-E_{\rm L}^{1-\alpha_{\rm ph}})}}.
\end{equation}
The $\gamma$-ray luminosity is calculated by
\begin{equation}
L_{\rm \gamma} = 4\pi d_{\rm L}^2(1+z)^{(\alpha_{\rm ph}-2)}F,
\end{equation}
where $d_{\rm L} = \frac{c}{H_{\rm 0}}\int^{1+z}_{1}\frac{1}{\sqrt{\Omega_{\rm m}x^{3}+1-\Omega_{\rm m}}}dx$ is a luminosity distance and $(1+z)^{(\alpha_{\rm ph}-2)}$ stands for a $K$-correction.

\begin{figure}[htbp]
\centering
\includegraphics[scale=0.7]{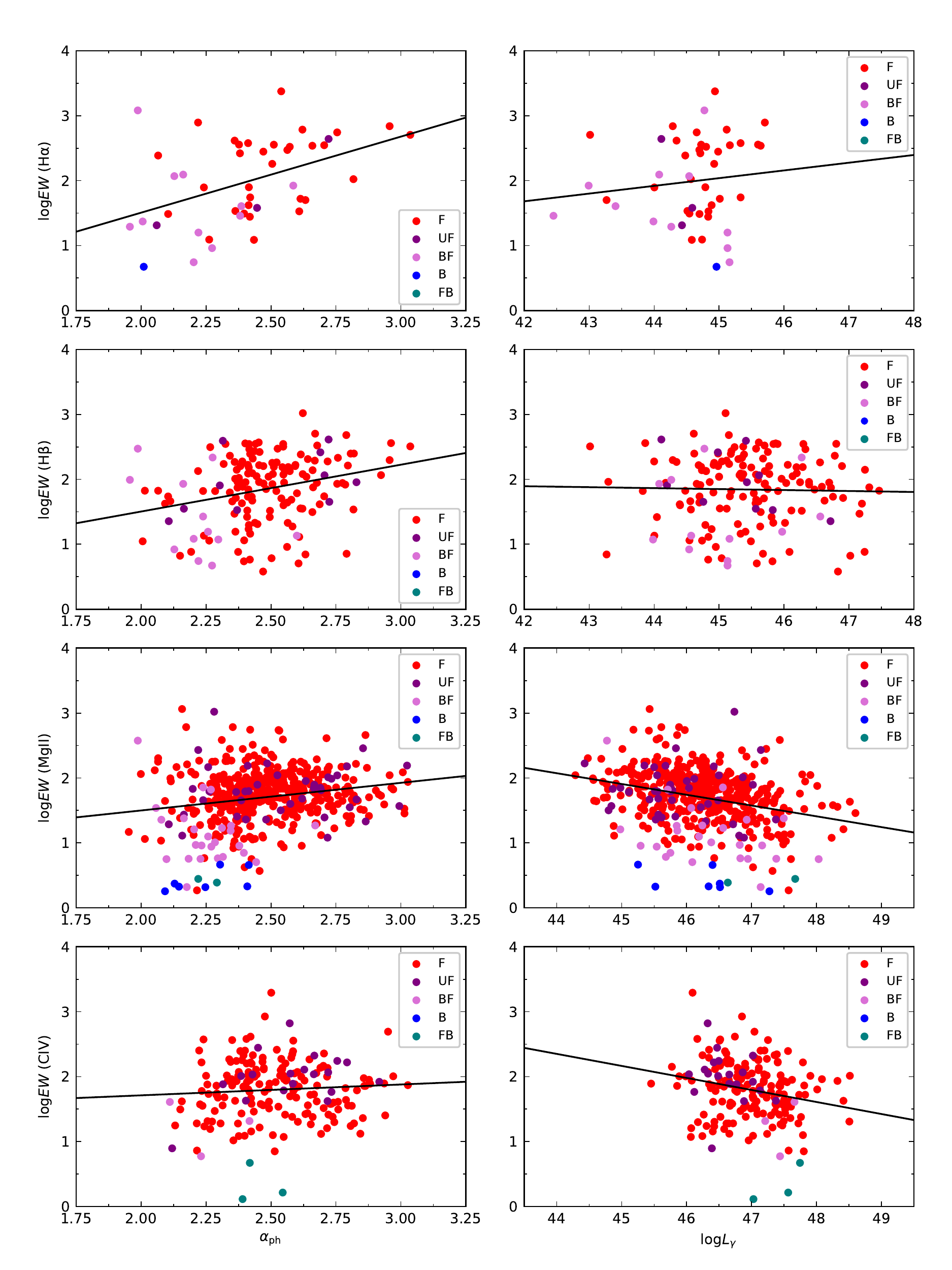}
\caption{The correlations between emission line ($\rm H{\alpha}$, $\rm H{\beta}$, $\rm Mg{II}$, and $\rm C{IV}$) $EW$ and GeV $\gamma$-ray parameters, photon spectral index $\alpha_{\rm ph}$ (left column) and $\gamma$-ray luminosity ${\rm log}L_{\gamma}$ (right column).
`F' stands for FSRQs, `UF' stands for the new confirmed FSRQs from BCUs in this work, `BF' stands for the `changing look' blazar that change from BL Lac to FSRQs, `B' stands for BL Lacs, `FB' stands for the `changing look' blazar that change from FSRQs to BL Lacs.}
\label{Fig_cor_g}
\end{figure}

Figure \ref{Fig_cor_g} shows the correlations between the $EW$ of emission lines ($\rm H{\alpha}$, $\rm H{\beta}$, $\rm Mg{II}$, and $\rm C{IV}$) and the GeV $\gamma$-ray parameters ($\gamma$-ray photon spectral index $\alpha_{\rm ph}$ and $\gamma$-ray luminosity $L_{\gamma}$).
The linear regression fitting results are shown in Table \ref{lin_cor_g}.

\section{Discussion}
\subsection{The completeness of this sample}
\begin{figure}[htbp]
\centering
\includegraphics[scale=0.75]{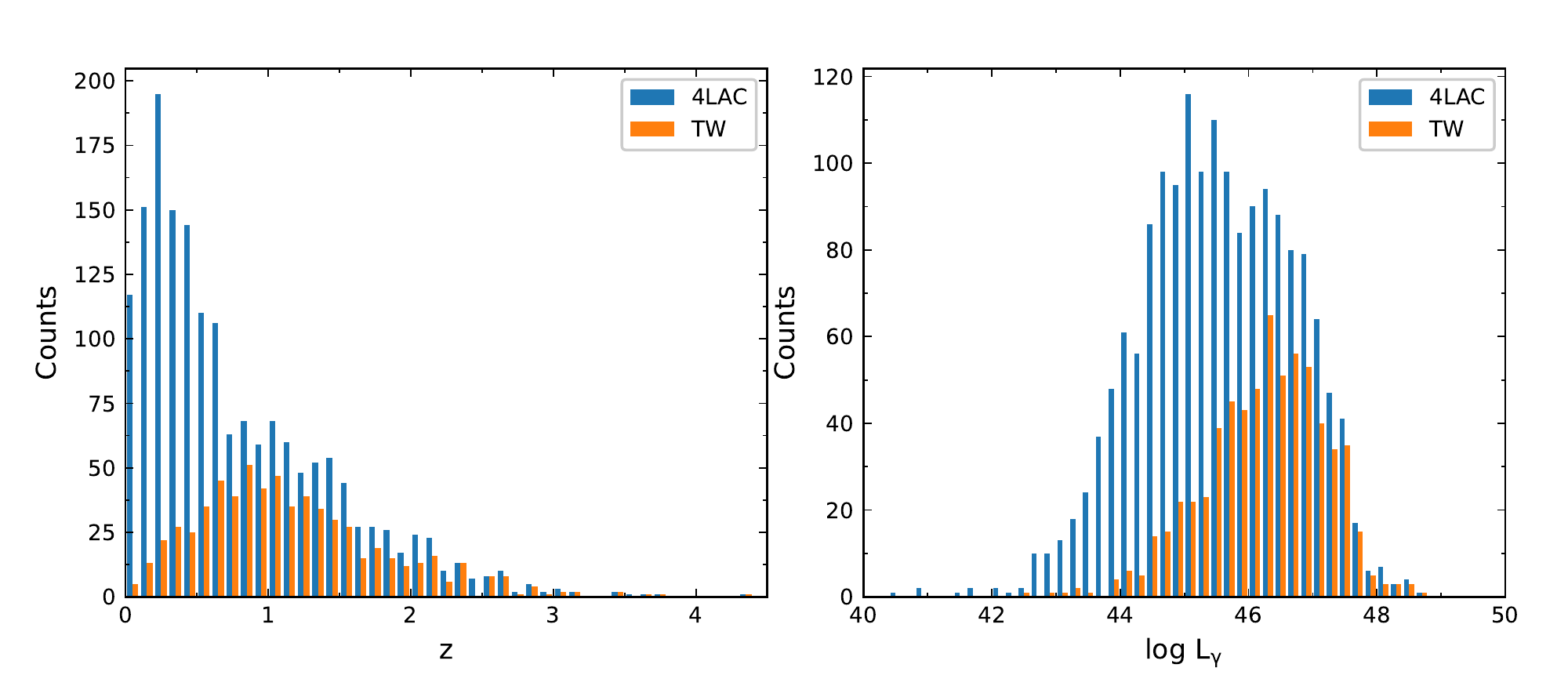}
\caption{The redshift and $\gamma$-ray luminosity distributions of the sources in our sample and in 4LAC\_DR2.
The blue bar stands for 4LAC sources, the orange bar stands for the sources in our sample.}
\label{Fig_hist}
\end{figure}

In this work, we have compiled a sample of blazars with emission line features from literature, these sources are all included in 4LAC\_DR2. 
To evaluate the completeness of our sample, we compare the redshift ($z$) and the $\gamma$-ray luminosity ($L_{\rm \gamma}$) from our sample and from the whole \textit{Fermi} blazar sample.
The redshift and $\gamma$-ray luminosity distributions for blazars in our sample and for 1701 blazars with known redshift in 4LAC\_DR2 are shown in Figure \ref{Fig_hist}.
The sources in our sample (this work, TW) have larger redshift and higher $\gamma$-ray luminosity, noting that most of the 4FGL\_DR2 blazars with $z < 1$ and ${\rm log}\, L_{\rm \gamma} < 46 \ {\rm erg \cdot s^{-1}}$ are not included in our sample.
The redshift ranges from 0.000017 (4FGL J0654.0-4152) to 4.313 (4FGL J1510.1+5702) with an average value of $\langle z^{\rm 4LAC} \rangle = 0.80 \pm 0.66$ for the 1701 4LAC\_DR2 blazars;
the redshift spans from 0.027 (4FGL J2204.3+0438) to 4.314 (4FGL J1510.1+5702) with an average value of $\langle z^{\rm TW} \rangle = 1.16 \pm 0.66$ for the 656 blazars in our sample.
The $\gamma$-ray luminosity ranges from 35.75 (4FGL J0654.0-4152) to 48.76 (4FGL J1833.6-2103) with an average value of $\langle {\rm log} \, L_{\rm \gamma}^{\rm 4LAC} \rangle = 45.49 \pm 1.27$ for the 1701 4LAC\_DR2 redshift-known blazars;
the $\gamma$-ray luminosity ranges from 42.45 (4FGL J2204.3+0438) to 48.60 (4FGL J1427.9-4206) with an average value of $\langle {\rm log} \, L_{\rm \gamma}^{\rm TW} \rangle = 46.22 \pm 0.94$ for the 656 blazars in our sample.
Anderson-Darling (A-D) test is applied to test if $z^{\rm TW}$ and $z^{\rm 4LAC}$, ${\rm log} \, L_{\rm \gamma}^{\rm TW}$ and ${\rm log} \, L_{\rm \gamma}^{\rm 4LAC}$ come from the same distributions.
The A-D test gives statistics 123.7 and 120.8 for redshift and $\gamma$-ray luminosity, respectively.
The values are both greater than the critical statistic of 6.5 for a significance level of 0.001 and this rejects the null hypothesis that two distributions come from the same distribution.
Thus, we can state that our sample is not a good representative of the whole \textit{Fermi} blazars, but a sample of brighter and more distant \textit{Fermi} blazars.
This sample incompleteness is caused by the selection criteria as we preferentially selected sources with emission line features in the optical spectrum, leading us to select the brighter and more distant \textit{Fermi} blazars. 
We caution the readers that our results are valid for this sample.

\subsection{`Changing-look' blazars}
\subsubsection{Comparing with `Changing-look' blazars in previous works}
AGNs are divided into `Type I' and `Type II' based on their optical spectra. 
The former displays a blue continuum from an accretion disk and broad emission lines created by photoionization, the latter shows only narrow lines and no continuum variability \citep{Khachikian1974, Peterson2004}.
The common understanding of these two categories is that the line of sight to the central engine is unobscured for Type I AGNs and obscured for Type II AGNs \citep{Antonucci1993, Urry1995}.
For the extreme AGNs, blazars are usually grouped into FSRQs and BL Lacs based on the $EW$.
However, these standard unification pictures for the difference between these classes meet challenges after the discovery of `Changing-look' AGNs or blazars.
The shifting between Type I AGNs and Type II AGNs was observed and reported \citep{Matt2003, Bianchi2005, Marchese2012, Shappee2014}, and the shifting was also reported between FSRQs and BL Lacs \citep{Isler2013, Isler2015, Mishra2021}.

There are works exploring `Changing-look' blazars.
\citet{Pena2021} carried out a sample of 26 `Changing-look' blazars by searching the available optical spectra in the Large Sky Area Multi-object Fiber Spectroscopic Telescope (LAMOST) Data Release 5 (DR5) archive \citep{Yao2019}.
\citet{Mishra2021} presented multi-wavelength photometric and spectroscopic monitoring observations of the blazar B2 1420+32, focusing on its outbursts in 2018–2020, and suggested that this source had transitioned between BL Lac and FSRQ states multiple times.

Cross-checking the sample in \citet{Pena2021} and \citealt{Mishra2021} with our sample, it is found that seven sources are confirmed as `Changing-look' blazars in common.
Three of them, 4FGL J1001.1+2911 (5BZB J1001+2911), 4FGL J1402.6+1600 (5BZB J1402+1559), and 4FGL J1503.5+4759 (TXS 1501+481) are indicated as `Changing-look' blazars in both our work and in \cite{Pena2021}.
The rest of the four sources, 4FGL J1043.2+2408 (5BZQ J1043+2408), 4FGL J1106.0.2+2813 (5BZQ J1106+2812), 4FGL J1321.1+2216 (5BZQ J1321+2216), and 4FGL J1422.3+3223 (B2 J1420+32) that are listed in Table \ref{tab_comm}, are classified as `Changing-look' blazars in \citet{Pena2021} and \citet{Mishra2021},  but they are not classified as `Changing-look' blazars in the present work because of different time domains of observed spectra.
4FGL J1043.2+2408, also known as 5BZQ J1043+2408, is contained in SDSS\_DR16 \citep{Ahumada2020} with a spectrum taken in 2013 March (MJD 56358) and was considered as an FSRQ due to a broad emission line of $\rm Mg{II}$; 
4FGL J1106.0+2813, also known as 5BZQ J1106+2812, was considered as an FSRQ in this work, \citet{Paliya2021} and 4LAC\_DR2 because its spectrum was taken from \citet{Shaw2012}, in which this source was classified as FSRQ;
4FGL J1321.1+2216, also known as 5BZQ J1321+2216, is contained in SDSS\_DR16 \citep{Ahumada2020} with a spectrum taken in 2012 May (MJD 56070), based on which we consider this source to be an FSRQ as also noted as FSRQ in 4LAC\_DR2.
However, these three sources show clear evidence of `Changing-look', from FSRQs to BL Lacs, when analyzing their latest spectra from LAMOST\_DR5, in which the spectra were taken in the duration of 2015 September to 2017 June.
4FGL J1422.3+3223, known as B2 1420+32, was classified as an FSRQ by both \citet{Paliya2021} and 4LAC\_DR2 based on the spectrum from SDSS\_DR16, which contains the spectrum data through 2018 August.
The new classification of BL Lac \citep{Mishra2021} makes this source be a ‘Changing-look’ blazars based on the spectroscopic study of outbursts in 2018–2020.

In total, we managed to obtain a sample of 56 `Changing-look' blazars, in which 52 (Table \ref{tab_cl}) are found in this work and 4 (Table \ref{tab_comm}) are collected from \citet{Pena2021} and \citet{Mishra2021}.
Among our 52 `Changing-look' blazars, there are 45 newly confirmed sources.

\subsubsection{The accretion ratio of `Changing-look' blazars}
The `Changing-look' AGNs/blazars were supposed to originate from obscuration of the quasar core by dusty clouds moving in the torus, but this explanation was basically dismissed because the expected high linear optical polarization was not widely observed \citep{Hutsemekers2019}.
An alternative explanation is that the `Changing-look' is arisen by a sudden change in accretion rate, broad emission lines emerge when the accretion rate increase and broad emission lines disappear when the accretion rate suddenly decreases.

The blazar classification has been studied in previous studies \citep{Ghisellini2011, Sbarrato2012, Xiong2014}, based on the normalized BLR luminosity ($L_{\rm BLR}/L_{\rm Edd}$) and the normalized $\gamma$-ray luminosity ($L_{\rm \gamma}/L_{\rm Edd}$, in Eddington units).
$L_{\rm BLR} = \xi L_{\rm Disk}$ and $L_{\rm Disk} = \eta \dot{M} c^{2}$, where $\xi$ is photoionization coefficient, $\eta$ is energy accretion efficiency, $\dot{M}$ is an accretion rate; 
$L_{\rm Edd} = \dot{M}_{\rm Edd} c^{2}$, where $\dot{M}_{\rm Edd}$ is an Eddington accretion rate.
Then one can get $\frac{L_{\rm BLR}}{L_{\rm Edd}} = \xi \eta \frac{\dot{M}}{\dot{M}_{\rm Edd}}$ by substituting $L_{\rm BLR}$ and $L_{\rm Edd}$.
\citet{Ghisellini2011} assumed $\dot{M}/\dot{M}_{\rm Edd} = 0.1$, together with the assumed $\xi = \eta = 0.1$, and suggested a separation value of $L_{\rm BLR}/L_{\rm Edd} = 1 \times 10^{-3}$ to separate FSRQs from BL Lacs, and FSRQs have higher $L_{\rm BLR}/L_{\rm Edd}$ than BL Lacs.
Later, \citet{Sbarrato2012} suggested a separation value of $5 \times 10^{-4}$ by using $\dot{M}/\dot{M}_{\rm Edd} = 0.05$.
\citet{Xiao2022} suggested that $\xi = \eta = 0.1$ may not appropriate for the blazars in our sample, and proposed the use of  $\xi = 0.11 $ and $\eta = 0.05$.
In this case, we can obtain $\left(L_{\rm BLR}/L_{\rm Edd}\right)^{\rm TW} = 5.5 \times 10^{-4}$ refer to $\dot{M}/\dot{M}_{\rm Edd} = 0.1$, or $\left(L_{\rm BLR}/L_{\rm Edd}\right)^{\rm TW} = 2.8 \times 10^{-4}$ refer to $\dot{M}/\dot{M}_{\rm Edd} = 0.05$.

We have collected $L_{\rm BLR} /L_{\rm Edd}$ from \citet{Xiao2022} for the 52 `Changing-look' blazars in this work and those 4 in \citet{Pena2021} and \citet{Mishra2021}, listed them for the 56 changing-look blazars in column (4) of Table \ref{tab_cl} and in column (7) of Table \ref{tab_comm}.
We notice that 14 (taking 100\%) `F$\to$B' and 38 (taking 90.5\%) `B$\to$F' blazars show $L_{\rm BLR}/L_{\rm Edd}$ greater than $2.8\times10^{-4}$, which means 92.9\% `Changing-look' blazars in our sample have values of $\dot{M}/\dot{M}_{\rm Edd}$ larger than 0.05;
and that 12 (taking 85.7\%) `F$\to$B' and 36 (taking 85.7\%) `B$\to$F' blazars show $L_{\rm BLR}/L_{\rm Edd}$ greater than $5.5\times10^{-4}$, which means 85.7\% `Changing-look' blazars in our sample have values of $\dot{M}/\dot{M}_{\rm Edd}$ larger than 0.1, see in Table \ref{tab_ratio}.
It is that the `Changing-look' blazars mostly lie above the dividing lines proposed by \citet{Ghisellini2011} and \citet{Sbarrato2012}.
\citet{Pei2022} studied the correlation $L_{\rm Disk} /L_{\rm Edd} \, {\rm vs} \, L_{\rm \gamma} /L_{\rm Edd}$, assuming $L_{\rm Disk}=10L_{\rm BLR}$, and it was clearly shown in Figure 5 of their work that the FSRQs and the `Changing-look' blazars have larger accretion ratio than the BL Lacs.
They have proposed an `appareling zone', $2.0\times10^{-4} \leq L_{\rm BLR}/L_{\rm Edd} \leq 8.5\times10^{-3}$, to select `Changing-look' blazars candidates, and we found that there are 46 (taking 82.1\%) `Changing-look' blazars in our sample that lie in this zone.
Thus, the `Changing-look' blazars have a larger accretion ratio than the normal BL Lacs, which are believed to have a lower accretion ratio.
And our result supports the explanation that the `Changing-look' originated from the sudden change in accretion rate.

\subsubsection{SED classification of `Changing-look' blazars}
Blazars are also divided into LSPs, ISPs, HSPs based on their synchrotron peak locations.
We notice that among the 56 `Changing-look' blazars, there are 54 sources with available SED classification from 4LAC\_R2, and 2 `B$\to$F' sources, 4FGL J0127.9+4857 and 4FGL J0823.3+2224, without SED classification.
For the 54 blazars with SED classification, all 14 `F$\to$B' blazars are associated to LSPs, 35 `B$\to$F' blazars are associated to LSPs, 4 `B$\to$F' blazars are associated to ISPs, 1 `B$\to$F' blazars are associated to HSPs.
It makes 90.7\% of `Changing-look' blazars are LSPs, and only a small fraction of ISPs or HSPs are associated to `Changing-look' blazars.
In this case, the LSPs could be a bank of `Changing-look' blazars, especially for the lower synchrotron peaked BL Lacs (LBLs).

\subsection{The correlations}
\subsubsection{The correlation between $EW$ and the GeV $\gamma$-ray parameters}
Figure \ref{Fig_cor_g} illustrates the correlation between the equivalent width of emission lines and the $\gamma$-ray photon index, and the observed $\gamma$-ray luminosity.
We note that there are positive correlations (with $p$ values less than 0.05), according to the fitting results in Table \ref{lin_cor_g}, between ${\rm log}\, EW$ and $\alpha_{\rm ph}$ for ${\rm H{\alpha}}$, ${\rm H{\beta}}$, and ${\rm Mg{II}}$, meanwhile, a positive trend (with $p$ values greater than 0.05) between ${\rm log} \, EW$ and $\alpha_{\rm ph}$ for ${\rm C}{\rm IV}$ is found.
The positive correlation and trend suggest that the stronger the emission line the softer the GeV spectrum.
There are anti-correlations between ${\rm log}\, EW$ and ${\rm log}\, L_{\rm \gamma}$ for ${\rm Mg}{\rm II}$ and ${\rm C}{\rm IV}$, but this anti-correlation is not found for either ${\rm H}{\rm \alpha}$ or ${\rm H}{\rm \beta}$.

\subsubsection{The Baldwin effect}
The correlation between the equivalent width and continuum luminosity is illustrated in Figure \ref{Fig_cor_line} and the corresponding regression results are tabulated in Table \ref{lin_cor_line}.
A trend of anti-correlation between ${\rm log}\, EW \, {\rm (H{\beta})}$ and ${\rm log}\, L_{\rm 5100}$ shows up; 
and solid anti-correlations between ${\rm log}\, EW \, {\rm (Mg{II})}$ and ${\rm log}\, L_{\rm 3000}$, between ${\rm log}\, EW \, {\rm (C{IV})}$ and ${\rm log}\, L_{\rm 1350}$ are found.
Our results of the anti-correlations are consistent with the results reported in many other previous works \citep{Dietrich2002, Shields2007, Kovacevic2010, Shemmer2015, Patino2016, Rakic2017}.

The anti-correlation between the broad-line equivalent width and the continuum luminosity of single-epoch observations of a large number of AGNs are known as the global `Baldwin effect' (hereafter, BEff) \citep{Baldwin1977, Carswell1978}.
The BEff indicates that the line flux is increasing more slowly than the local continuum (or is constant) because $EW$ is the ratio of line flux to the local continuum flux.
The BEff is well established for broad emission lines in the UV/optical regions \citep{Shields2007, Dietrich2002}, even for narrow lines \citep{Dietrich2002, Kovacevic2010}, and it is found that it steepens with increasing ionization potential \citep{Zheng1993}.
Various mechanisms have been proposed as possible interpretations of the BEff, such as a luminosity-dependent ionization continuum and the BLR covering factor \citep{Mushotzky1984, Zheng1993}, the geometrical effect of an inclination-dependent anisotropic continuum \citep{Netzer1985}, different variability patterns in the thermal and non-thermal components of the continuum \citep{Kinney1990, Patino2016}, or processes that involve different Eddington ratio or black hole mass \citep{Xu2008, Bian2012}, etc.
The most widely accepted explanation is that the ionization continuum softens as the luminosity increases \citep{Zheng1993}, so that high-luminosity AGNs decrease the fraction of ionizing photons for broad emission line formation.

In this work, we found a slope of $-0.08\pm0.06$ for the correlation between $EW$ of ${\rm H{\beta}}$ emission line and continuum luminosity at $5100\, \mathring{\rm A}$, however, a chance probability of 0.23 suggests the correlation is not evident.
\citet{Rakic2017} obtained a slope of -0.0467 of $EW\,{\rm (H{\beta})}$ against $L_{\rm 5100}$ in a logarithmic diagram, and claimed that no evidence of significant BEff for ${\rm H{\beta}}$ emerged.
Both of the results, in this work and in \citet{Rakic2017}, agree with previous findings, which concluded that no BEff was present in the broad Balmer lines \citep{Dietrich2002, Kovacevic2010}).
But, the BEff is found for ${\rm Mg{II}}$ and ${\rm C{IV}}$ emission lines with slopes of $-0.24\pm0.03$ and of $-0.25\pm0.05$ (see Table \ref{lin_cor_line}) in this work.
Our results are consistent with the results of BEff for ${\rm Mg{II}}$ and ${\rm C{IV}}$ emission lines in \citet{Patino2016}, in which they found the BEff for these two lines and reported slopes of 0.20 and 0.21 were derived, respectively.

It is found that there is a difference in BEff anti-correlation between blazars and radio-quiet (RQ) AGNs, the former shows a steeper \textit{anti-correlation} than the latter one \citet{Patino2016}.
We suggest the steepen of BEff for blazars results from the IC scattering of broad emission line photons from BLR for two reasons.
On one hand, consider the case that the continuum luminosity is identical for blazars and RQs, the $EW$ for blazars should be smaller than it is for RQs because a significant number of broad-emission-line photons are fed to the IC process, and scattered to $\gamma$-ray band, resulting in a high Compton dominance \citep{Abdo2010, Ghisellini2011, Paliya2021} and a weaker emission line;
On the other hand, when the continuum luminosity gets stronger, the bolometric luminosity should increase.
The bolometric luminosity is dominated by the IC emission (or the $\gamma$-ray luminosity, \citealp{Ghisellini2010, Ghisellini2014, Xiong2014, Xiao2022}), therefore, increased bolometric luminosity yields more external photons (e.g. broad-emission-line photons) to feed the IC process.
Then the emission lines of blazars (mainly FSRQs) get weaker and the difference on $EW$ become larger with the increase of continuum luminosity compare to the RQs.
Consequently, a steepening of BEff is formed for blazars.

In this scenario, it is natural that strong emission lines should be able to contribute more seed photons to the IC process, therefore, we can expect a positive correlation between $EW$ and IC luminosity.

We collected the flux of the IC peak ($F_{\rm IC}$) from \citet{Paliya2021} to calculate the IC peak luminosity ($L_{\rm IC}=4\pi d_{\rm L}^{2} F_{\rm IC}$, in the observer frame).
However, the observed IC luminosity is boosted by a Doppler beaming effect \citep{Dermer1995, Paliya2015, Paliya2021} and gives the intrinsic IC luminosity (in the source rest frame) as
$L^{\rm in}_{\rm IC} = L_{\rm  IC}/\delta^{4}$,
where $\delta$ is the Doppler factor.
Doppler factors are available in different literature and are given in discrepancy.
In this work, we employ the method that was proposed by \citet{Zhang2020}, in which they proposed to use $\gamma$-ray luminosity and broad line region luminosity ($L_{\rm BLR}$) to calculate $\delta$ of $\gamma$-ray emission, to calculate $\delta$ for our sources.
And the data of $L_{\rm BLR}$ is calculated based on the work \citet{Paliya2021} and \citet{Xiao2022}.

\begin{figure}[htbp]
\centering
\includegraphics[scale=0.9]{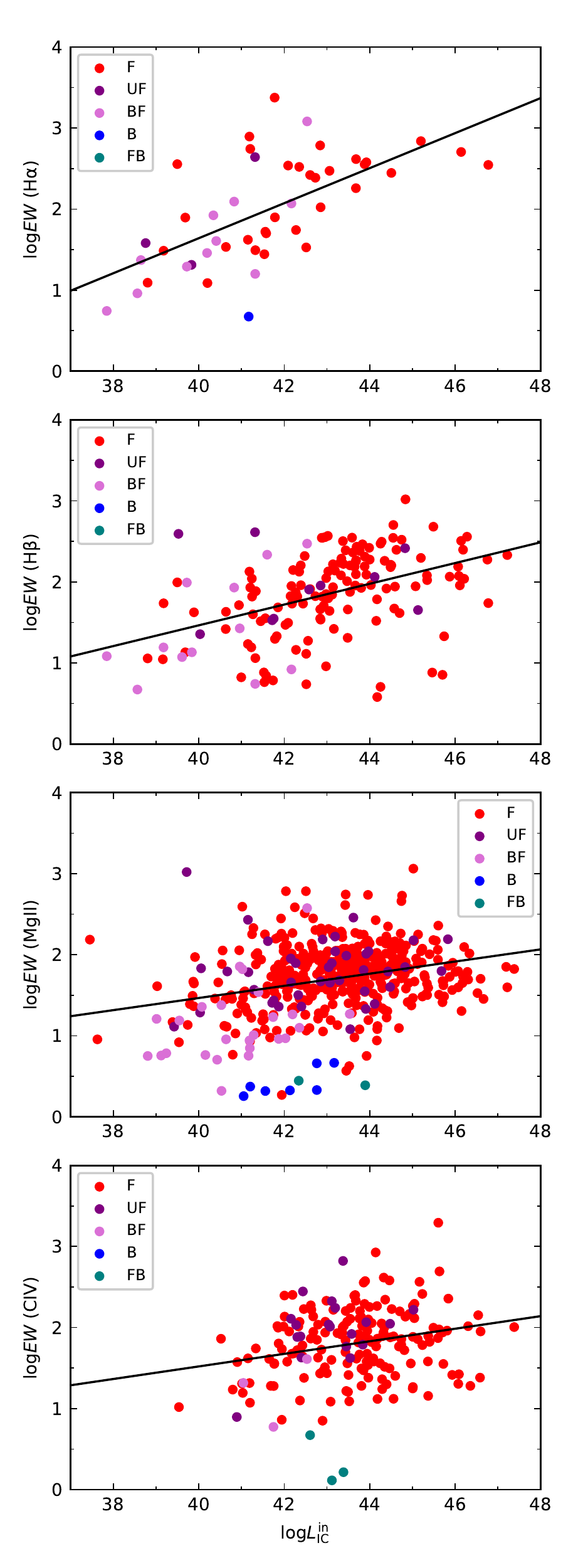}
\caption{The correlations between emission line ($\rm H{\alpha}$, $\rm H{\beta}$, $\rm Mg{II}$, and $\rm C{IV}$) $EW$ and the intrinsic IC luminosity (${\rm log}L_{\rm IC}^{\rm in}$).
`F' stands for FSRQs, `UF' stands for the new confirmed FSRQs from BCUs in this work, `BF' stands for the `changing look' blazar that change from BL Lac to FSRQs, `B' stands for BL Lacs, `FB' stands for the `changing look' blazar that change from FSRQs to BL Lacs.}
\label{Fig_cor_IC_in}
\end{figure}

The results between $EW$ and intrinsic inverse Compton luminosity are shown in Table \ref{lin_cor_IC_in} and in Figure \ref{Fig_cor_IC_in}.
The results illustrate that the $EW$ is positively correlated with intrinsic inverse Compton luminosity, suggesting that sources with the stronger intrinsic inverse Compton luminosity tend to have stronger emission lines.
Moreover, these results have confirmed our prediction, a positive correlation between $EW$ and inverse Compton luminosity, and have proven that the steepening of BEff for blazars (mostly FSRQs) is, indeed, caused by the IC scattering of broad-emission-line photons.

\section{Conclusion}
In this work, we aim to study the emission line property and its connection with the non-thermal emission of blazars, we collected the equivalent width of emission lines, $\gamma$-ray emission and SED information for a sample of 656 \textit{Fermi} blazars.
We have studied the blazar classification according to the $EW$ and the correlation between the $EW$ and $\gamma$-ray luminosity, and the global `Baldwin effect'.

Our main results are as follows:
(1) There are, out of the 656 \textit{Fermi} blazars, 55 previously classified as BCUs are now classified as FSRQs;
(2) We find 52 `Changing-look' blazars through the study of the $EW$, among them there is 10 FSRQs change to BL Lacs and 42 BL Lacs change to FSRQs.
Besides, 45 of them are newly confirmed as `Changing-look' blazars;
(3) The accretion rate ($\dot{M}/\dot{M}_{\rm Edd}$) of the 52 `Changing-look' blazars are calculated.
We notice that there are 92.9\% `Changing-look' blazars in our sample with $\dot{M}/\dot{M}_{\rm Edd} >0.05$ and 85.7\% `Changing-look' blazars in our sample with $\dot{M}/\dot{M}_{\rm Edd} >0.1$, suggesting the `Changing-look' blazars have larger accretion ratio.
Besides, we notice that 90.7\% of the `Changing-look' blazars in this work are LSPs and suggest the LSPs are a bank of `Changing-look' blazars;
(4) The global `Baldwin effect' is confirmed for blazars (mostly FSRQs) in this work, the results indicate a steeper anti-correlation of equivalent width (${\rm log}\, EW$) against continuum luminosity (${\rm log}\, L_{\rm 3000}$ and ${\rm log}\, L_{\rm 1350}$) than that of the radio-quiet (RQ) AGNs;
(5) We propose that the steepening of global BEff is caused by the inverse Compton scattering of broad-emission-line photons and predict a positive correlation between the equivalent width (${\rm log}\, EW$) and the inverse Compton luminosity (${\rm log}\, L_{\rm IC}$).
This prediction is indeed correct and the $EW$ of H$\alpha$, H$\beta$, MgII and CIV are positively correlated with the intrinsic inverse Compton peak luminosity.

\begin{acknowledgments}
We thank the support from our laboratory, the key laboratory for astrophysics of Shanghai, we thank Dr. Vaidehi S. Paliya for his kindly help and for sharing data.
Meanwhile, L. P, Fu acknowledges the support from the National Natural Science Foundation of China (NSFC) grants 11933002, STCSM grants 18590780100, 19590780100, SMEC Innovation Program 2019-01-07-00-02-E00032 and Shuguang Program 19SG41.
S. H, Zhang acknowledges the support from by Natural Science Foundation of Shanghai (20ZR1473600).
J. H, Fan acknowledges the support by the NSFC (NSFC U2031201, NSFC 11733001).
\end{acknowledgments}

\begin{table}[htbp]\scriptsize
\centering
\caption{Optical, $\gamma$-ray, and SED parameters of 656 4LAC blazars}
\label{tab_og}
\begin{tabular}{lcccccccccc} 
\hline
4FGL name & z & $EW_{\rm H{\alpha}}$ & $EW_{\rm H{\beta}}$ & $EW_{\rm MgII}$ & $EW_{\rm CIV}$ &${\rm log}L_{\rm BLR}$ & ${\rm log}F_{\rm IC}$  & ${\rm f}_{\gamma}$ & $\alpha_{\rm ph}$ & Class \\
(1) & (2) & (3) & (4) & (5) & (6) & (7) & (8) & (9) & (10) & (11) \\ 
\hline
J0001.5+2113	&	0.439	&	344.8$\pm$12.91	    &	31.29$\pm$3.91	    &	37.2$\pm$2.74	&	                   &  43.74  &    -10.48    &   1.36E-09    &	2.66	&	F \\
J0004.3+4614	&	1.81	&				        &				        &				    &	246.4$\pm$17.19	   &  45.07  &    -11.7     &   2.41E-10	&	2.58	&	F \\
J0004.4-4737	&	0.88	&				        &				        &	22.54$\pm$5.12	&				       &  44.10  &    -11.24    &   4.36E-10	&	2.37	&	F \\
J0006.3-0620	&	0.347	&	117.39$\pm$26.82    &	8.33$\pm$5.75	    &				    &				       &  43.60  &    -12.04    &   1.40E-10	&	2.13	&	B \\
J0010.6+2043	&	0.598	&				        &   172.38$\pm$18.94    &	112.67$\pm$4.28	&				       &  44.34  &    -11.97    &   1.73E-10	&	2.32	&	F \\ 
\hline
\end{tabular}
\tablecomments{Column definitions: 
(1) 4FGL name;
(2) redshift;
(3) equivalent width of $\rm H{\alpha}$ emission line in units of $\mathring{\rm A}$;
(4) equivalent width of $\rm H{\beta}$ emission line in units of $\mathring{\rm A}$;
(5) equivalent width of $\rm Mg{II}$ emission line in units of $\mathring{\rm A}$;
(6) equivalent width of $\rm C{IV}$ emission line in units of $\mathring{\rm A}$;
(7) luminosity of the broad emission line region;
(8) flux of inverse Compton peak; 
(9) integral photon flux from 1 to 100 GeV, in units of ${\rm photon \cdot cm^{-2} \cdot s^{-1}}$; 
(10) photon index; 
(11) 4LAC\_DR2 Classification, `B' stands for BL Lacs, `F' stands for FSRQs, `U' stands for BCUs;
Note that features in column (2)-(6), and (8) are obtained from \citet{Paliya2021}, features in column (9)-(11) are obtained from \citet{Ajello2020}.  
Only 5 objects are presented here, the table is available in its entirety in machine-readable form.}
\end{table}

\begin{table}[htbp]\scriptsize
\centering
\caption{The new classification for BCUs in our sample}
\label{tab_u}
\begin{tabular}{lccc} 
\hline 
\hline
4FGL name & z & Class & New classification  \\
(1) & (2) & (3) & (4)  \\ 
\hline
J0014.3-0500	&	0.791	&	U	&	F	\\
J0030.6-0212	&	1.804	&	U	&	F	\\
J0036.9+1832	&	1.595	&	U	&	F	\\
J0040.9+3203	&	0.632	&	U	&	F	\\
J0143.5-3156	&	0.374	&	U	&	F	\\
J0204.8+1513	&	0.407	&	U	&	F	\\
J0223.5-0928	&	1.005	&	U	&	F	\\
J0226.3-1845	&	1.67	&	U	&	F	\\
J0327.5-1805	&	0.73	&	U	&	F	\\
J0430.2-0356	&	0.628	&	U	&	F	\\
J0516.8-0509	&	1.417	&	U	&	F	\\
J0621.2-4648	&	1.212	&	U	&	F	\\
J0622.9+3326	&	1.062	&	U	&	F	\\
J0658.1-5840	&	0.421	&	U	&	F	\\
J0725.8-0054	&	0.128	&	U	&	F	\\
J0728.0+6735	&	0.844	&	U	&	F	\\
J0749.3+4453	&	0.559	&	U	&	F	\\
J0821.1+1007	&	0.954	&	U	&	F	\\
J0904.0+2724	&	1.721	&	U	&	F	\\
J0904.9-5734	&	0.697	&	U	&	F	\\
J0941.7+4125	&	0.816	&	U	&	F	\\
J0943.7+6137	&	0.791	&	U	&	F	\\
J0949.7+5819	&	1.424	&	U	&	F	\\
J1017.8+0715	&	1.54	&	U	&	F	\\
J1018.9+1043	&	0.66	&	U	&	F	\\
J1047.9+0055	&	0.252	&	U	&	F	\\
J1054.2+3926	&	2.635	&	U	&	F	\\
J1124.4+2308	&	0.795	&	U	&	F	\\
J1129.2-0529	&	0.922	&	U	&	F	\\
J1131.4-0504	&	0.263	&	U	&	F	\\
J1139.0+4033	&	2.361	&	U	&	F	\\
J1159.2-2227	&	0.565	&	U	&	F	\\
J1205.8+3321	&	1.007	&	U	&	F	\\
J1243.0+3950	&	1.22	&	U	&	F	\\
J1248.9+4840	&	1.856	&	U	&	F	\\
J1319.5-0045	&	0.891	&	U	&	F	\\
J1323.0+2941	&	1.142	&	U	&	F	\\
J1329.4-0530	&	0.576	&	U	&	F	\\
J1412.9+5018	&	1.53	&	U	&	F	\\
J1418.4+3543	&	0.825	&	U	&	F	\\
J1454.0+4927	&	2.106	&	U	&	F	\\
J1615.6+2130	&	1.627	&	U	&	F	\\
J1627.3+4758	&	2.32	&	U	&	F	\\
J1720.2+3824	&	0.454	&	U	&	F	\\
J1821.6+6819	&	1.69	&	U	&	F	\\
J2136.2-0642	&	0.941	&	U	&	F	\\
J2140.5-6731	&	2.009	&	U	&	F	\\
J2211.2-1325	&	0.392	&	U	&	F	\\
J2253.3+3233	&	0.257	&	U	&	F	\\
J2311.7+2604	&	1.748	&	U	&	F	\\
J2313.9-4501	&	2.877	&	U	&	F	\\
J2318.2+1915	&	2.163	&	U	&	F	\\
J2326.2+0113	&	1.6	    &	U	&	F	\\
J2339.6+0242	&	2.661	&	U	&	F	\\
J2352.9+3031	&	0.876	&	U	&	F	\\ 
\hline
\end{tabular}
\tablecomments{Column definitions: 
(1) 4FGL name;
(2) redshift;
(3) 4LAC\_DR2 classification, `U' denotes BCU;
(4) new classification, `F' denotes FSRQ;}
\end{table}

\begin{table}[htbp]\scriptsize
\centering
\caption{The `Changing-look' blazars in our sample}
\label{tab_cl}
\begin{tabular}{lcccc} 
\hline 
\hline
4FGL name & z & Change & $L_{\rm BLR}/L_{\rm Edd}$ & SED classification \\
(1) & (2) & (3) & (4) & (5) \\ 
\hline
J0102.8+5824	&	0.644	&	F $\to$ B	&	0.00034	&	LSP	\\
J0337.8-1157	&	3.442	&	F $\to$ B	&	0.00776	&	LSP	\\
J0347.0+4844	&	2.039	&	F $\to$ B	&	0.00035	&	LSP	\\
J0521.3-1734	&	0.347	&	F $\to$ B	&	0.00377	&	LSP	\\
J0539.6+1432	&	2.71	&	F $\to$ B	&	0.01659	&	LSP	\\
J0539.9-2839	&	3.104	&	F $\to$ B	&	0.00380	&	LSP	\\
J0601.1-7035	&	2.4	    &	F $\to$ B	&	0.00071	&	LSP	\\
J1816.9-4942	&	1.7	    &	F $\to$ B	&	0.01213	&	LSP	\\
J2015.5+3710	&	0.855	&	F $\to$ B	&	0.00734	&	LSP	\\
J2121.0+1901	&	2.04	&	F $\to$ B	&	0.00222	&	LSP	\\
J0006.3-0620	&	0.347	&	B $\to$ F	&	0.00097	&	LSP	\\
J0127.9+4857	&	0.065	&	B $\to$ F	&	0.00350	&		\\
J0203.7+3042	&	0.761	&	B $\to$ F	&	0.02678	&	LSP	\\
J0209.9+7229	&	0.895	&	B $\to$ F	&	0.00230	&	LSP	\\
J0238.6+1637	&	0.94	&	B $\to$ F	&	0.00240	&	LSP	\\
J0334.2-4008	&	1.359	&	B $\to$ F	&	0.04245	&	LSP	\\
J0407.5+0741	&	1.139	&	B $\to$ F	&	0.00097	&	LSP	\\
J0428.6-3756	&	1.11	&	B $\to$ F	&	0.00598	&	LSP	\\
J0433.6+2905	&	0.91	&	B $\to$ F	&	0.00228	&	LSP	\\
J0438.9-4521	&	2.027	&	B $\to$ F	&	0.00065	&	LSP	\\
J0516.7-6207	&	1.3	    &	B $\to$ F	&	0.00021	&	LSP	\\
J0538.8-4405	&	0.896	&	B $\to$ F	&	0.00561	&	LSP	\\
J0629.3-1959	&	1.718	&	B $\to$ F	&	0.00172	&	LSP	\\
J0654.7+4246	&	0.129	&	B $\to$ F	&	0.00108	&	LSP	\\
J0710.9+4733	&	1.292	&	B $\to$ F	&	0.00194	&	LSP	\\
J0814.4+2941	&	0.374	&	B $\to$ F	&	0.00092	&	HSP	\\
J0823.3+2224	&	0.951	&	B $\to$ F	&	0.00575	&		\\
J0831.8+0429	&	0.174	&	B $\to$ F	&	0.00064	&	LSP	\\
J0832.4+4912	&	0.548	&	B $\to$ F	&	0.00260	&	LSP	\\
J1001.1+2911	&	0.556	&	B $\to$ F	&	0.00125	&	LSP	\\
J1031.1+7442	&	0.123	&	B $\to$ F	&	0.00515	&	ISP	\\
J1058.0+4305	&	1.308	&	B $\to$ F	&	0.00355	&	LSP	\\
J1058.4+0133	&	0.892	&	B $\to$ F	&	0.00177	&	LSP	\\
J1058.6-8003	&	0.581	&	B $\to$ F	&	0.00038	&	LSP	\\
J1147.0-3812	&	1.053	&	B $\to$ F	&	0.00168	&	LSP	\\
J1250.6+0217	&	0.954	&	B $\to$ F	&	0.00918	&	LSP	\\
J1331.2-1325	&	0.251	&	B $\to$ F	&	0.00212	&	LSP	\\
J1402.6+1600	&	0.245	&	B $\to$ F	&	0.00014	&	ISP	\\
J1412.1+7427	&	0.436	&	B $\to$ F	&	0.00091	&	ISP	\\
J1503.5+4759	&	0.345	&	B $\to$ F	&	0.00398	&	LSP	\\
J1647.5+4950	&	0.047	&	B $\to$ F	&	0.00148	&	LSP	\\
J1751.5+0938	&	0.322	&	B $\to$ F	&	0.00703	&	LSP	\\
J1800.6+7828	&	0.691	&	B $\to$ F	&	0.00186	&	LSP	\\
J1806.8+6949	&	0.05	&	B $\to$ F	&	0.00741	&	LSP	\\
J1954.6-1122	&	0.683	&	B $\to$ F	&	0.00347	&	LSP	\\
J2134.2-0154	&	1.283	&	B $\to$ F	&	0.00087	&	LSP	\\
J2152.5+1737	&	0.872	&	B $\to$ F	&	0.00046	&	LSP	\\
J2202.7+4216	&	0.069	&	B $\to$ F	&	0.00020	&	LSP	\\
J2204.3+0438	&	0.027	&	B $\to$ F	&	0.00020	&	ISP	\\
J2216.9+2421	&	1.033	&	B $\to$ F	&	0.00289	&	LSP	\\
J2315.6-5018	&	0.811	&	B $\to$ F	&	0.00143	&	LSP	\\
J2357.4-0152	&	0.816	&	B $\to$ F	&	0.00277	&	LSP	\\
\hline
\end{tabular}
\tablecomments{Column definitions: 
(1) 4FGL name;
(2) redshift;
(3) blazar type change, `B' denotes BL Lac, `F' denotes FSRQ;;
(4) normalized BLR luminosity;
(5) SED classification, `LSP' denotes low synchrotron peak sources, `ISP' denotes intermediate synchrotron peak sources, and `HSP' denotes high synchrotron peak sources.}
\end{table}

\begin{table}[htbp]
\centering
\setlength{\tabcolsep}{16pt}
\renewcommand{\arraystretch}{1.3}
\caption{The correlation between $EW$ and continuum luminosity}
\label{lin_cor_line}
\begin{tabular}{lccccc}
\hline
\hline
y vs x & $(a \pm \Delta a)$ & $(b \pm \Delta b)$ & $N$ & $r$ & $p$\\
\hline
${\rm log}\, EW \ {\rm (H{\beta})}\ vs \ {\rm log}\, L_{\rm 5100}$ & -0.08 $\pm$ 0.06 & 5.23 $\pm$ 2.84 & 156 & -0.1 & 0.23 \\
${\rm log}\, EW \ {\rm (Mg{II})}\ vs \ {\rm log}\, L_{\rm 3000}$ & -0.24 $\pm$ 0.03 & 12.74 $\pm$ 1.29 & 481 & -0.36 & $1.4 \times 10^{-16}$ \\
${\rm log}\, EW \ {\rm (C{IV})}\ vs \ {\rm log}\, L_{\rm 1350}$ & -0.25 $\pm$ 0.05 & 13.26 $\pm$ 2.41 & 191 & -0.33 & $3.9 \times 10^{-6}$ \\
\hline
\end{tabular}
\end{table}

\begin{table}[htbp]
\centering
\setlength{\tabcolsep}{16pt}
\renewcommand{\arraystretch}{1.3}
\caption{The correlation between $EW$ and GeV $\gamma$-ray parameters}
\label{lin_cor_g}
\begin{tabular}{lccccc}
\hline
\hline
y vs x & $(a \pm \Delta a)$ & $(b \pm \Delta b)$ & $N$ & $r$ & $p$\\
\hline
${\rm log}\, EW \ {\rm (H{\alpha})}\ vs \ \alpha_{\rm ph}$ & 1.17 $\pm$ 0.35 & -0.83 $\pm$ 0.85 & 47 & 0.45 & 0.002 \\
${\rm log}\, EW \ {\rm (H{\alpha})}\ vs \ {\rm log}\, L_{\rm \gamma}$ & 0.12 $\pm$ 0.14 & -3.31 $\pm$ 6.39 & 47 & 0.12 & 0.41 \\
${\rm log}\, EW \ {\rm (H{\beta})}\ vs \ \alpha_{\rm ph}$ & 0.72 $\pm$ 0.20 & 0.06 $\pm$ 0.50 & 156 & 0.28 & $5 \times 10^{-4}$ \\
${\rm log}\, EW \ {\rm (H{\beta})}\ vs \ {\rm log}\, L_{\rm \gamma}$ & -0.01 $\pm$ 0.05 & 2.51 $\pm$ 2.22 & 156 & -0.02 & 0.76 \\
${\rm log}\, EW \ {\rm (Mg{II})}\ vs \ \alpha_{\rm ph}$ & 0.43 $\pm$ 0.10 & 0.65 $\pm$ 0.24 & 481 & 0.20 & $1.6 \times 10^{-5}$ \\
${\rm log}\, EW \ {\rm (Mg{II})}\ vs \ {\rm log}\, L_{\rm \gamma}$ & -0.17 $\pm$ 0.03 & 9.38 $\pm$ 1.17 & 481 & -0.29 & $1.5 \times 10^{-10}$ \\
${\rm log}\, EW \ {\rm (C{IV})}\ vs \ \alpha_{\rm ph}$ & 0.17 $\pm$ 0.17 & 1.38 $\pm$ 0.43 & 191 & 0.07 & 0.33 \\
${\rm log}\, EW \ {\rm (C{IV})}\ vs \ {\rm log}\, L_{\rm \gamma}$ & -0.19 $\pm$ 0.06 & 10.50 $\pm$ 2.82 & 191 & -0.22 & 0.002 \\
\hline
\end{tabular}
\end{table}

\begin{table}[htbp]
\centering
\setlength{\tabcolsep}{16pt}
\renewcommand{\arraystretch}{1.3}
\caption{The overlapped `Changing-look' blazars with previous works}
\label{tab_comm}
\begin{tabular}{lcccccc}
\hline
\hline
4FGL name & Other name & P21 & M21 & TW & Change & $L_{\rm BLR}/L_{\rm Edd}$    \\
(1) & (2) & (3) & (4) & (5) & (6) & (7) \\
\hline
J1043.2+2408	&	5BZQ J1043+2408	    &	Y	&		&	N	&   F$\to$B     &   0.00132 \\
J1106.0+2813	&	5BZQ J1106+2812 	&	Y	&		&	N	&   F$\to$B     &   0.00554 \\
J1321.1+2216	&	5BZQ J1321+2216 	&	Y	&		&	N	&   F$\to$B     &   0.01110 \\
J1422.3+3223	&	B2 1420+32	        &		&	Y	&	N	&   F$\to$B     &   0.01010 \\
\hline
\end{tabular}
\tablecomments{P21: \citet{Pena2021}; M21: \citet{Mishra2021}; `Y' denotes `Yes' and `N' denotes `No'.}
\end{table}

\begin{table}[htbp]
\centering
\setlength{\tabcolsep}{12pt}
\renewcommand{\arraystretch}{1.3}
\caption{The $L_{\rm BLR} /L_{\rm Edd}$ for `Changing-look' blazars in this work}
\label{tab_ratio}
\begin{tabular}{lcccc}
\hline
\hline
Change type & $>5.5\times10^{-4}$ & $<5.5\times10^{-4}$ & $>2.8\times10^{-4}$ & $<2.8\times10^{-4}$ \\ \hline
F $\to$ B & 12 & 2 & 14 & 0 \\ \hline
B $\to$ F & 36 & 6 & 38 & 4 \\ \hline
\end{tabular}
\end{table}

\begin{table}[htbp]
\centering
\setlength{\tabcolsep}{16pt}
\renewcommand{\arraystretch}{1.3}
\caption{The correlations between EW and the intrinsic inverse Compton peak luminosity}
\label{lin_cor_IC_in}
\begin{tabular}{lccccc}
\hline
\hline
y vs x & $(a \pm \Delta a)$ & $(b \pm \Delta b)$ & $N$ & $r$ & $p$\\
\hline
${\rm log} EW \ {\rm (H{\alpha})}\ vs \ {\rm log} L_{\rm IC}^{\rm in}$ & 0.22 $\pm$ 0.04 & -7.01 $\pm$ 1.60 & 47 & 0.64  & $1.1 \times 10^{-6}$ \\
${\rm log} EW \ {\rm (H{\beta})} \ vs \ {\rm log} L_{\rm IC}^{\rm in}$ & 0.13 $\pm$ 0.02 & -3.66 $\pm$ 0.89 & 156 & 0.44 & $6.0 \times 10^{-9}$ \\
${\rm log} EW \ {\rm (Mg{II})}   \ vs \ {\rm log} L_{\rm IC}^{\rm in}$ & 0.07 $\pm$ 0.01 & -1.53 $\pm$ 0.52 & 481 & 0.27 & $9.1 \times 10^{-10}$ \\
${\rm log} EW \ {\rm (C{IV})}    \ vs \ {\rm log} L_{\rm IC}^{\rm in}$ & 0.08 $\pm$ 0.02 & -1.58 $\pm$ 1.03 & 191 & 0.23 & 0.001 \\
\hline
\end{tabular}
\end{table}

\bibliography{lib}{}
\bibliographystyle{aasjournal}

\end{document}